\documentstyle[prl,aps,12pt]{revtex}

\newcommand{\two}[2]{#2}   
\newcommand{\E}{\varepsilon}
\newcommand{\ts}{
   \parbox{1mm}{
      \setlength{\unitlength}{1mm}
      \begin{picture}(1,10)
         \thinlines
         \put(0.5,-0.2){\line(0,1){7}}
      \end{picture}
   }
}
\newcommand{\lts}{
   \parbox{1mm}{
      \setlength{\unitlength}{1mm}
      \begin{picture}(1,12)
         \thinlines
         \put(0.5,-0.2){\line(0,1){11}}
      \end{picture}
   }
}

\newcommand{\be}{\begin{equation}}
\newcommand{\ee}{\end{equation}}
\newcommand{\bea}{\begin{eqnarray}}
\newcommand{\eea}{\end{eqnarray}}
\newcommand{\nn}{\nonumber}

\newcommand{\lineup}{\raisebox{5mm}[0mm][0mm]{\makebox[0mm][l]{\epsfxsize=8.9cm
\epsfbox{./linel.eps}}}}
\newcommand{\linedown}{\raisebox{5mm}[0mm][0mm]{\makebox[0mm][l]{\hspace{9.1cm}
\epsfxsize=8.9cm\epsfbox{./liner.eps}}}}
\newcommand {\eq}[1]{(\ref{#1})}
\newcommand {\p}{\partial}
\newcommand{\half}{\frac{1}{2}}

\title{Critical Discussion of the 2-Loop Calculations for the KPZ-Equation}
\author{\centerline{Kay J\"org Wiese}
\centerline{Fachbereich Physik, Universit\"at GH Essen,  45117 Essen,
Germany}
}
\date{June 13, 1997}
\begin{document}
\maketitle
\begin{abstract}\noindent
In this article, we perform a careful analysis of the renormalization
procedure used in existing calculations to derive critical
exponents for the KPZ-equation at
2-loop order. This analysis explains the discrepancies between the results
of the different groups. The correct critical exponents in $d=2+\E$
dimensions  at the crossover between weak- and strong-coupling regime
are
$\chi={\cal O}(\E^3)$ and $z=2+{\cal O}(\E^3)$.
No strong-coupling fixed point exists at 2-loop order.
\end{abstract}

\two{\begin{multicols}{2}\narrowtext}{}
\tableofcontents
\noindent
\section{Introduction}
The Kardar-Parisi-Zhang equation \cite{KPZ}
\bea
\frac{\p h(x,t)}{\p t} & =& \nu \nabla^2 h(x,t) + \frac\lambda 2 \left( \nabla
h(x,t)\right)^2 + \eta(x,t) \label{KPZ} \\
\overline{\eta(x,t) \eta(x',t')} & =& 2D \delta^d(x-x') \delta(t-t')
\eea
plays a central role
as simplest field-theoretic model for non-linear growth and has
extensively been studied  during the last years. Whereas
the situation is clear for space-dimension $d=1$,
the (physical more interesting) case of $d\ge2$
can  only be attacked by approximative
methods or field-theoretic perturbative
expansions. Using the latter, the fixed point
structure of the renormalization group flow for $d=2+\E$ has been obtained.
Two domains can be distinguished:
For small effective coupling
\be
        g=\frac{\lambda^2 D}{\nu^3} \ ,
\ee
the renormalization group flow tends to
0 in the long-wavelength limit, for large coupling to a
strong coupling fixed point $g=g_{\mbox\scriptsize{sc}}$.
The crossover takes place at $g=g_{\mbox\scriptsize{co}}$,
which turns out to be  of order $\E$ in an $\E$-expansion
and can therefore be studied perturbatively.

Three such calculations have been performed, one by Sun and Plischke
\cite{SunPlischke94}, another by {Teodorovich} \cite{Teodorovich96}, the
third by Frey and T\"auber \cite{FreyTaeuber94}.  The results are
contradictory. We were able to trace back the difference to a problem
in the application of the renormalization group procedure, which
lead \cite{SunPlischke94} and \cite{Teodorovich96} to incorrect
results, see the discussion in sections \ref{Sun},
\ref{Teodorovich} and \ref{Frey}. We will first derive some elementary and
hopefully well-known properties of the renormalization group, which will help
us to make the point clear.

\section{Some elementary properties about renormalization and
the freedom to choose a renormalization scheme}
\label{Element prop}
In the following, we want to exploit the freedom one has to choose
renormalization factors in the dimensional regularization scheme.
As an example, take any massless theory and
suppose that the derivative with respect to $k^2$ of
the bare and renormalized 2-point functions (we denote this by a
dot) at a given renormalization
scale $\mu$ are related by
\be \label{def Z}
\dot \Gamma^{(2)}_{B}
= Z(g_R) \,
\dot\Gamma^{(2)}_{R}
\ .
\ee
The factor $Z(g_R)$, which is a function of the renormalized coupling $g_R$ and
$\E$ only, is introduced to define a finite renormalized two-point
function  $\Gamma^{(2)}_{R}$.  Then, it is standard to obtain the
anomalous dimension of the field, given by the
renormalization group $\zeta$-function, see e.g.\ \cite{Zinn},
as variation of $\ln(Z)$ with respect to the renormalization scale
$\mu$, keeping bare quantities fixed
\bea \label{zeta(g)}
\zeta(g_R) &=& -\half \mu \frac{\partial}{\partial \mu}\lts_B \ln Z(g_R)\nn\\
        &=& -\half \beta(g_R) \frac{\partial}{\partial g_R} \ln Z(g_R) \ .
\eea
In the second line, we have introduced the renormalization group
$\beta$-function, which is obtained from the variation of the
renormalized coupling $g_R$ with respect to the
renormalization scale $\mu$:
\be
\beta(g_R) = \mu\frac{\partial}{\partial \mu}\lts_B g_R
\ee
and where renormalized and bare coupling constants $g_R$ and $g_B$
are related by\be
g_B = \mu^{\E} Z_g^{-1} (g_R) g_R \ .
\ee
The critical point $g^*$ is the zero of the $\beta$-function
\be
\beta(g^*)=0 \ .
\ee
The critical exponent $\zeta^{*}$ is
\be
\zeta^* = \zeta( g^* ) \ .
\ee
All these definitions are standard \cite{Zinn} and are used (in slightly
different
notations) in \cite{SunPlischke94}, \cite{Teodorovich96} and
\cite{FreyTaeuber94}. (\cite{Teodorovich96} uses a cut-off
regularization, but there is no principal  difference.)
We now ask the question: Is the renormalization factor $Z(g_R)$, which was
 introduced in
equation \eq{def Z}, unique? The answer is no. $Z(g_R)$ can be multiplied by 
any
finite factor  $Z_f(g_R)$ and $\Gamma^{(2)}_R$ rests finite.
Instead of $Z(g_R)$ one could therefore use $Z'(g_R)$,
\be
Z'(g_R)=Z(g_R) \times Z_f (g_R)\ .
\ee
Let us verify that this is consistent and that the critical exponent
$\zeta^*$ is unchanged. The new $\zeta'$-function is
\bea
\zeta'(g_R) &=& -\half \mu \frac{\partial}{\partial \mu}\ln Z'(g_R) \nn\\
&=&\zeta(g_R) -\half \beta(g_R) \frac{\partial}{\partial g_R} \ln Z_f(g_R)  \ .
\eea
As $Z_f(g_R)$ is finite for any $g_R$, $\beta(g_R) \frac{\partial}{\partial  
g_R} \ln Z_f(g_R)$ tends to 0 for $g_R\to g^*$ and we obtain the
result
\be
        \zeta'(g^*) =\zeta(g^*) \ .
\ee
One can also prove that the freedom to choose $g_R$ is a simple
finite reparametrization
\be
g_R'=g_R \times Z_f(g_R) \ .
\ee
One has however to be careful to make this change of variables in
every renormalization factor.

Let us now discuss the minimal subtraction scheme.
Suppose that $\Gamma^{(2)}_{B}$
at the renormalization point $\mu$ satisfies up to 2-loop order,
i.e.\ up to order $g_R^2$
\two{\end{multicols}\widetext\noindent\lineup}{}
\be \label{lang}
\dot \Gamma^{(2)}_{B} = \left(
1+ a_{-1}  \frac {{g_R}} \E  + a_0 {g_R} +a_1 {g_R}\E + \ldots +
b_{-2}  \frac {g_R^2} {\E^2} + b_{-1}  \frac {g_R^2} {\E}
+ b_{0}  {g_R^2}  + b_{1}   {g_R^2} \E + \ldots
\right) 1 \ee
where $a_i$ and $b_i$ are constants and where
we have suppressed higher order terms in $\E$. We now
want to introduce renormalization factors. Of course, setting in \eq{def Z}
\be
Z(g_R)=1+ a_{-1}  \frac {{g_R}} \E  + a_0 {g_R} +a_1 {g_R}\E + \ldots +
b_{-2}  \frac {g_R^2} {\E^2} + b_{-1}  \frac {g_R^2} {\E}
+ b_{0}  {g_R^2}  + b_{1}   {g_R^2} \E + \ldots
\ee
\two{\begin{multicols}{2}\narrowtext\noindent\linedown}{}would render
$\Gamma^{(2)}_R$ finite. (The dots stand for the
same higher order terms as in \eq{lang}.)
But this is not the simplest
possible choice. The idea of the minimal subtraction scheme
is now, to choose a $Z$-factor, which contains only the pole-terms
in $\E$. One is therefore tempted to replace $Z$ by
\be
Z_{\mbox{\scriptsize PM}}(g_R) = 1+ a_{-1}  \frac {{g_R}} \E +
b_{-2}  \frac {g_R^2} {\E^2} + b_{-1}  \frac {g_R^2} {\E} \ .
\ee
The reader is invited to stop for a second to think about this choice
before going on to read.

Let us see whether $Z_{\mbox{\scriptsize PM}}$ correctly
renormalizes the theory. Then, the ratio of $Z$ and $Z_{\mbox{\scriptsize PM}}$
has to be finite. Expanding up to order $g_R^2$ and dropping all terms of
order $\E^0$ and higher, we obtain:
\be
        \frac{Z(g_R)}{Z_{\mbox{\scriptsize PM}}(g_R)} = 1 -  a_0 a_{-1}
\frac{g_R^2}{\E}
        + \ldots
\ .
\ee
This ratio is not finite and $Z_{\mbox{\scriptsize PM}}$
therefore not a correct renormalization factor! We will henceforth call
this wrong prescription to find the $Z$-factor in the minimal subtraction
scheme ``pseudo minimal-subtraction scheme'' in order to make clear
that this very special mistake was done.

 The correct factor to be used in a minimal
subtraction scheme would be
\be
Z_{\mbox{\scriptsize MS}}(g_R) = 1+ a_{-1}  \frac {{g_R}} \E +
b_{-2}  \frac {g_R^2} {\E^2} + \left(b_{-1}-a_0a_{-1}\right)  \frac {g_R^2}
{\E}
\ee
as the reader may easily convince himself by the construction given above.

Apparently, the term $a_0$ cannot be neglected at 2-loop order.
Let us determine more generally, which terms may intervene in a $n$-loop
calculation.
First of all, the perturbative fixed point in  an $\E$-expansion is of order
$\E$
\be \label{gsimE}
        g^* \sim \E \ .
\ee
(There are situations where $g^*$ is of order $\E^2$ or higher. For
those cases the following argument has to be modified.)
Let us now determine in which order of $\E$ a term of order
\be
\frac{g_R^k}{\E^l}
\ee
in $Z(g_R)$ will contribute to the $\zeta$-function at $g_R=g^*$.
Taking care of the factor $\E$, which comes from
the variation of $\ln Z(g_R)$ in the definition of $\zeta(g_R)$,
equation \eq{zeta(g)},  the answer is with the help of \eq{gsimE}
\be
\E\, \frac{(g^*)^k}{\E^l}= \E^{1+k-l} \ .
\ee
To calculate the critical exponent $\zeta^*$
up to order $\E^n$, one therefore has to determine
 all contributions to $Z$ which are proportional
to $g_R^k$ up to order $\E^{n-k-1}$.

This means that to calculate up to order $\E^2$, one can drop terms
 from  $Z$ and retain only
\be
Z_{\mbox{\scriptsize d}}(g_R)=
1+ a_{-1}  \frac {{g_R}} \E  + a_0 {g_R} +
b_{-2}  \frac {g_R^2} {\E^2} + b_{-1}  \frac {g_R^2} {\E}
\ .
\ee
Note again that $a_0$ cannot be dropped.

In the next sections we will apply these considerations to the
2-loop calculations of the KPZ-equation in the literature.

\section{Analysis of the work by Sun and Plischke}
\label{Sun}
In this section we are going to analyse the 2-loop renormalization group
calculations by Sun and Plischke \protect\cite{SunPlischke94}, see
also the comment in \cite{commentSun95}.
{F}rom their Appendixes A and B it is clear that the finite part of
the 1-loop contribution is dropped. They write: ``In this
Appendix we list all expressions of Feynman graphs \ldots
and their final results in $2-\E$ dimensions to order $O(\E^{-1})$."
As there is a finite contribution at 1-loop order, the discussion given
above shows that their results
are incorrect.
We would have liked to incorporate the missing term in their calculations
but as not all $Z$-factors are given explicitly we have to leave it to the
authors.

\section{Analysis of the work by Teodorovich}
\label{Teodorovich}
Let us also look at the article by Teodorovich \cite{Teodorovich96}.
Teodorovich is conscious that the finite part of the 1-loop diagrams
enters as insertion of these diagrams into 2-loop diagrams and he
explicitly calculates the finite contributions to the 1-loop diagrams in
his equations (5.1) and (5.2). But then he also
 drops the finite contribution to $\Sigma^{R}$,
which is still present in equation (5.1),
in equation (5.3). We are therefore led to the same conclusion as in
the case of  Sun and Plischke:
The results are incorrect. Let us however mention that Teodorovich
 does not agree with our criticism. He states to have used the BPHZ
 procedure and inserted 1-loop renormalized terms in order to
calculate the renormalization factors \cite{Teodorovich private}.
It is however impossible from his article \cite{Teodorovich96}
to get the precise procedure enabling one to verify his claim.

\section{Analysis of the work by Frey and T\"auber}
\label{Frey}
In this section we are going to analyse the 2-loop renormalization group
calculations by Frey and T\"auber \cite{FreyTaeuber94}, see also
their comment in \cite{repcommentFreyT94}.

They introduce the following two renormalization constants for the
renormalization of $D$ and $\nu$, see
\cite{FreyTaeuber94} equation (3.43) and (3.44)
\two{\end{multicols}\widetext\noindent\lineup}{}
\bea
Z_D &=& 1-\frac {\hat g_B} \E - (d-1) \frac{ \hat g_B^2} \E
        -\frac{d-2}{d} \frac{ \hat g_B^2}{2\E}
        +(d-1) \frac{ \hat g_B^2} {\E^2} +{\cal O}(\hat g_B^3)
\label{ZD}
\\
Z_\nu&=& 1+\frac{d-2}d \left[ \frac{\hat g_B}{\E} +(d-1) \frac{\hat g_B^2}{2\E}
+\frac{d-2}{d}\frac{\hat g_B^2}{2\E} -(d-1) \frac{\hat g_B^2}{2\E^2}\right]
-\frac{d-1}{d} \frac{\hat g_B^2}{2\E}+{\cal O}(\hat g_B^3) \ .
\label{Znu}
\eea
\two{\begin{multicols}{2}\narrowtext\noindent\linedown}{}The notation is
\bea
        \hat g_B&=& \mu^\E g_B \\
  \E &=& d-2
\label{eps=d-2}
\ .
\eea
(Note that our $\hat g_B= \hat g_0$ in the notation of \cite{FreyTaeuber94}.
To simplify the considerations, we also have replaced the function $F_\nu (d)$
in (3.44) of \cite{FreyTaeuber94} by its value at $d=2$, see (A38) of
\cite{FreyTaeuber94}. The difference is a term of order $\E^0$ and can be
neglected in 2-loop order.)   As we will check later these $Z$-factors
  subtract all divergences correctly. Let us first explain the appearance
of factors $d$
and $\E$. Frey and T\"auber \cite{FreyTaeuber94} use a dimensional
regularization prescription, where divergences appear as poles in $1/\E$. This
is the origin of the factors $1/\E$ and $1/\E^2$ in \eq{ZD} and \eq{Znu}. On
the other hand, there are integrals of the type
\be
\int_k \left(2(\vec k \vec p)^2 -k^2 p^2\right) f(k^2)
\ee
which are identical to
\be
p^2\frac{2-d}d \int_k k^2 f(k^2)
\ .
\ee
Frey and T\"auber call the factors $(d-2)$ ``geometrical''  because they
result from scalar products in the integrand. The renormalization scheme in
Ref. \cite{FreyTaeuber94} is such that those geometrical terms are not $\E$
expanded in order to meet the condition that a fluctuation-dissipation theorem
is valid in $d=1$ dimensions. This procedure results in the
appearance of the first non-trivial term in $Z_\nu$ (see equation \eq{Znu}),
\be \label{1}
\frac{d-2}d  \frac{\hat g_B}{\E} .
\ee
From the reasoning in \cite{FreyTaeuber94} one could  gain the impression
that such a term can be dropped, if the distinction between factors $d-2$ and
$\E$ is not made. This would be incorrect, since -- as we have
explained in detail in section \ref{Element prop} -- this term has to be
retained in order to renormalize the theory.
We would write \eq1 as
\be \label{2}
\left( \half  + {\cal O}( \E) \right) {\hat g_B}
\ee
and using our findings at the end of section \ref{Element prop} conclude that
the term of order $ \E^0$ has to be taken into account.  (To be precise: In
section \ref{Element prop} we had argued on the level of the $Z$-factors as
functions of the renormalized coupling $g_R$. As $g_B=g_R+{\cal O} (g_R^2)$
this does not affect our reasoning for the one-loop term in \eq{1} and \eq{2}.)

This analysis shows that the distinction between factors $d-2$ and $\E$ is not
necessary in order to renormalize the theory above $d=2$ dimensions.
Especially, one may not misinterpret \cite{FreyTaeuber94} in the way that
one could choose $d$ fixed  in the socalled ``geometrical''
  factors and then perform the limit $\E \to 0$ elsewhere. This would
result in a divergent term of order $1/\E$ in the renormalization group
functions, violating the very principle of renormalizability.

The correct procedure to follow is the following: Expand the
$Z$-factors in $\E$ only. Retain terms of order $\hat g_B$, $\hat g_B/\E$,
$\hat g_B^2/\E$ and $\hat g_B^2/\E^2$. (Of course, there would be no harm in
keeping higher order terms in $\E$, but this is unnecessary.)  Then calculate
the renormalization group functions.

We refrain from discussing the validity of a perturbative renormalization
group approach below the lower critical dimension $d=2$.
Such a scheme would have to meet certain criteria, such as the
validity of the fluctuation-dissipation theorem in $d=1$. But, it is not
clear to us how this procedure should be defined.

In order to demonstrate the correctness of our procedure and in order to
show that  factors $d-2$ and
$\E$ have not to be distinguished, we use the $Z$ factors of
\cite{FreyTaeuber94} in our scheme to establish the critical exponents.
In the view of our discussion in section \ref{Element prop} we first check
that the finite term in the factor $Z_D$ is correctly taken into
account. This is trivially true
 as normalizations in \cite{FreyTaeuber94} are chosen such that the only
integral
which has to be calculated in 1-loop order gives exactly $1/\E$ and {\em no}
finite contribution.

Replacing $d=2+\E$ and retaining only terms
of order $\hat g_B$, $\hat g_B/\E$, $\hat g_B^2/\E$ and $\hat g_B^2/\E^2$,
the $Z$-factors are:
\bea
{ Z_D(\hat g_B)} &=&  1 -
{\displaystyle \frac {{ \hat g_B}}{{ \E}}} + {\displaystyle
\frac {{ \hat g_B}^{2}}{{ \E}^{2}}}
\label{ZDred}
\\
{ Z_{\nu}(\hat g_B)} &=&  1 +
{\displaystyle \frac {1}{2}}{ \hat g_B} - {\displaystyle \frac {1
}{2}}{\displaystyle \frac {{ \hat g_B}^{2}}{{ \E}}}
\label{Znured}\\
{ Z_g(\hat g_B)} &=&  1 -  \left(
   {\displaystyle \frac {3}{2}} + {\displaystyle \frac {1
}{{ \E}}}   \right) { \hat g_B} +  \left(
{\displaystyle \frac {3}{{ \E}}}
 + {\displaystyle \frac {1}{{ \E}^{2}}}   \right)
{ \hat g_B}^{2}
\label{Zgred}
\ .
\eea
The last factor, $Z_g$, was calculated via the relation
\be
Z_g = \frac{Z_D}{Z_{\nu}^3}
\ee
as stated in \cite{FreyTaeuber94}.
It is now necessary to transform to renormalized
quantities. Replacing $\hat g_B$ in \eq{ZDred}, \eq{Znured} and \eq{Zgred}
by (note that this is sufficient in 2-loop order)
\be
\hat g_B = g_R +\left(\frac 1\E + \frac32 \right) g_R^2 + {\cal O}(g_R^3)
\ee
we obtain
\bea
{ Z_D(g_R)} &=&  1 -
{\displaystyle \frac {g_R}{{ \E}}} - {\displaystyle
\frac {3}{2}}{\displaystyle \frac {{ g_R}^{2}}{{ \E}}}
\\
{ Z_{\nu}(g_R)} &=& 1 +
{\displaystyle \frac {1}{2}}{ g_R}  \\
{ Z_g(g_R)} &=&  1 - \frac32 g_R -\frac{g_R}\E
\ .
\eea
The renormalization-group functions as defined in
\cite{FreyTaeuber94} are
\bea
\zeta_D(g_R) &=& \mu \frac{\partial }{\partial \mu}\lts_B \ln Z_D(g_R) \nn\\
        &=& \beta(g_R) \frac{\partial}{\partial g_R} \ln Z_D(g_R) \\
\zeta_{\nu}(g_R) &=& \mu \frac{\partial }{\partial \mu}\lts_B \ln
Z_{\nu}(g_R) \nn\\
        &=& \beta(g_R) \frac{\partial}{\partial g_R} \ln Z_{\nu}(g_R) \\
\beta(g_R) &=& \mu \frac{\partial }{\partial \mu}\lts_B g_R \nn\\
        &=& \frac{\E g_R}{1-g_R\frac{\partial}{\partial g_R} \ln Z_g(g_R)}
\nn\\
        &=& g_R\left( \E +\zeta_D(g_R) -3 \zeta_{\nu}(g_R)  \right)
\ .
\eea
Using our results from above, we obtain
the following 2-loop renormalization-group functions
\bea
\zeta_D(g_R) &=& -g_R -\frac32 g_R^2  \\
\zeta_{\nu}(g_R) &=& \frac \E 2 g_R -  \half  g_R^2 \\
\beta(g_R) &=& \E g_R -\left(1+\frac32 \E \right) g_R^2   \label{perbeta} \ ,
\eea
which are identical to the results obtained in equations (3.62)-(3.64)
of \cite{FreyTaeuber94} after a full $\E$ expansion.
There is one non-trivial ($g^*\ne0$)  zero of the $\beta$-function. It
is located at
\be
g^*=\E- \frac32 \E^2 +{\cal O}(\E^3)
\ .
\ee
Note that no fixed point of order 1 (``non-perturbative fixed point'')
exists in contrast to  \cite{SunPlischke94} and \cite{Teodorovich96}.
This yields the critical exponents
\bea
\zeta_D^* &=& -\E+{\cal O }(\E^3) \\
\zeta_{\nu}^* &=&  {\cal O }(\E^3)
\ .
\eea
The more standard quantities, the
 roughness-exponent $\chi$ and the dynamical exponent $z$ are
\bea
\chi&=& {\cal O }(\E^3) \label{finzeta}\\
z &=& 2+{\cal O }(\E^3)
\label{z}
\ .
\eea
In addition, the correction to scaling exponent
\be
\omega = \frac{\partial }{\partial g_R} \beta(g_R)\ts_{g_R=g^*}
\ee
is easily read off from \eq{perbeta}
\be
\omega = \E+ {\cal O }(\E^3)\ .
\ee
We have given all calculations in this explicit form to convince
the reader that the standard $\E$-expansion is sufficient in order
to obtain the critical exponents given in \eq{finzeta} and \eq{z}.

For all fans of the minimal subtraction scheme,
we leave it as an exercise that via a variable-transformation
and finite renormalizations one can transform to the following
$Z$-factors in the minimal-subtraction scheme
\bea
Z_D(g_R) &=& 1-\frac{g_R}\E +{\cal O}(g_R^3) \nn\\
Z_{\nu}(g_R) &=& 1 +{\cal O}(g_R^3)\\
Z_g(g_R) &=& 1-\frac{g_R}\E +{\cal O}(g_R^3)\nn
\ .
\eea
Using these $Z$-factors, one of course obtains the same critical exponents
to order $\E^2$.

We also found some minor misprints in \cite{FreyTaeuber94}. For the readers
conveniance we give a list here \cite{misprints}.

\section{Conclusions}
In this article we have analyzed the 2-loop renormalization group calculations
by Sun and Plischke \cite{SunPlischke94}, by {Teodorovich}
\cite{Teodorovich96} and by Frey and T\"auber \cite{FreyTaeuber94}.  Sun and
Plischke \cite{SunPlischke94} and {Teodorovich} \cite{Teodorovich96} don't
correctly take into account the finite part of the 1-loop diagrams. Therefore,
their results are incorrect.

The renormalization scheme by Frey and T\"auber \cite{FreyTaeuber94} leads
to the correct result. From the reasoning given in their paper one could,
however, gain the wrong impression that a distinction between socalled
geometrical factors and $1/\E$ poles is necessary in order to get a valid
renormalization of the KPZ equation. We have shown here, that such a
distinction is in fact not necessary above the lower critical dimension
$d=2$. This conclusion is supplemented
through work by L\"assig, who showed \cite{Laessig95}, using the mapping to
directed polymers, that the predicted critical exponents

\bea
\zeta&=&0 \nn \\
z &=& 2
\eea
are correct to all orders in $\E$.
This result had already been obtained in a different context in
\cite{ImbrieSpencer1988}.

On the technical level, we have shown that the dimensional regularization
scheme works perfectly well and has not to be modified in order to calculate
the critical exponents for the KPZ-equation to order $\E^2$  above the
lower critical dimension.  The analysis of the critical behavior below
$d=2$ is more subtle and it is presently not clear whether there is a valid
concept for the analysis of the strong coupling behavior.

\acknowledgements
I thank H.W.\ Diehl, M.\ Kardar, J.\ Krug, L.\ Sch\"afer and especially
J.\ Hager for valuable discussions.


\two{\end{multicols}}{}
\end{document}